\documentclass[preprint]{emulateapj}

\newcommand{\out}[1]{}  

\newcommand{\ngc}{\hbox{NGC\,891}}

\newcommand{\qts}{\hbox{${\rm quanta}\,{\rm s}^{-1}$}}
\newcommand{\cms}{\hbox{${\rm cm^{-2}}$}}

\newcommand{\gcc}{\hbox{${\rm g\,cm^{-3}}$}}
\newcommand{\cmc}{\hbox{${\rm cm^{-3}}$}}
\newcommand{\kms}{\hbox{${\rm km\,s^{-1}}$}}

\newcommand{\flusur}{\hbox{${\rm erg}\,{\rm cm}^{-2}\,{\rm s}^{-1}\,{\rm arcsec}^{-2}$}}

\newcommand{\ergs}{\hbox{${\rm erg\, {\rm s}^{-1}}$}}

\newcommand{\Ap}{\hbox{$A^{\prime}$}}
\newcommand{\Mw}{\hbox{$M_{w}$}}
\newcommand{\co}{\hbox{$c_{o}$}}
\newcommand{\cl}{\hbox{$c_{l}$}}
\newcommand{\effha}{\hbox{$\eta$}}
\newcommand{\rhol}{\hbox{$\rho_{l}$}}

\newcommand{\vw}{\hbox{$V_{w}$}}
\newcommand{\vt}{\hbox{$v_{t}$}}

\newcommand{\no}{\hbox{$n_{o}$}}
\newcommand{\nou}{\hbox{$n_{-1}$}}

\newcommand{\Tnii}{\hbox{$T_{\rm NII}$}}
\newcommand{\Tw}{\hbox{$T_{w}$}}
\newcommand{\Tx}{\hbox{$T_{X}$}}

\newcommand{\To}{\hbox{$T_{o}$}}
\newcommand{\barT}{\hbox{$\bar T$}}

\newcommand{\Po}{\hbox{$P_{o}$}}

\newcommand{\gam}{\hbox{$\Gamma$}}
\newcommand{\lo}{\hbox{$l_{o}$}}

\newcommand{\Uo}{\hbox{$U_o$}}
\newcommand{\Uf}{\hbox{$U_{-5}$}}

\newcommand{\phiw}{\hbox{$\varphi_{w}$}}
\newcommand{\phix}{\hbox{$\varphi_{X}$}}
\newcommand{\phis}{\hbox{$\varphi_{\ast}$}}

\newcommand{\Nx}{\hbox{$N_{\rm X}$}}
\newcommand{\No}{\hbox{$N_{\rm o}$}}

\newcommand{\QH}{\hbox{$Q_{\rm H}$}}
\newcommand{\QHe}{\hbox{$Q_{\rm He}$}}

\newcommand{\Lha}{\hbox{L$_{\rm H\alpha}$}}
\newcommand{\Fha}{\hbox{F$_{\rm H\alpha}$}}
\newcommand{\map}{\hbox{{\sc mappings i}c}}
\newcommand{\sed}{\hbox{{\sc sed}}}

\newcommand{\hb}{H$\beta$}

\newcommand{\ha}{\hbox{H$\alpha$}}

\newcommand{\hei}{\hbox{He\,{\sc i}}}
\newcommand{\heiw}{\hbox{He\,{\sc i}\,$\lambda $5876}}
\newcommand{\neii}{\hbox{[Ne\,{\sc ii}]}}
\newcommand{\neiiw}{\hbox{[Ne\,{\sc ii}]\,12.81\,$\mu$m}}
\newcommand{\neiii}{\hbox{[Ne\,{\sc iii}]}}
\newcommand{\neiiiw}{\hbox{[Ne\,{\sc iii}]\,15.56\,$\mu$m}}

\newcommand{\siiiw}{\hbox{[S\,{\sc iii}]\,18.71\,$\mu$m}}
\newcommand{\sii}{\hbox{[S\,{\sc ii}]}}
\newcommand{\siir}{\hbox{[S\,{\sc ii}]$\lambda $6716}}

\newcommand{\nii}{\hbox{[N\,{\sc ii}]}}
\newcommand{\niiw}{\hbox{[N\,{\sc ii}]$\lambda $6583}}
\newcommand{\niitw}{\hbox{[N\,{\sc ii}]$\lambda $5755}}

\newcommand{\oiii}{\hbox{[O\,{\sc iii}]}}

\newcommand{\oiiitw}{\hbox{[O\,{\sc iii}]$\lambda $4363}}
\newcommand{\oii}{\hbox{[O\,{\sc ii}]}}

\newcommand{\oi}{\hbox{[O\,{\sc i}]}}
\newcommand{\oiw}{\hbox{[O\,{\sc i}]$\lambda $6300}}

\newcommand{\hi}{\hbox{H\,{\sc i}}}
\newcommand{\hii}{\hbox{H\,{\sc ii}}}

\shorttitle{Mixing layers and DIGs} \shortauthors{Binette et al.}

\begin{document}


\title{Photoionized mixing layer models of the diffuse ionized gas}


\author{Luc Binette\altaffilmark{1,2}, Nahiely Flores-Fajardo\altaffilmark{2},
Alejandro C. Raga\altaffilmark{3}, Laurent Drissen\altaffilmark{1}}



\author{Christophe Morisset\altaffilmark{2}}


\altaffiltext{1}{D\'{e}partement de physique, de g\'{e}nie physique et
d'optique \& Centre de recherche en astrophysique du Qu\'ebec,
Universit\'{e} Laval, Qu\'{e}bec, Qc,
G1V\,0A6.} \altaffiltext{2}{Instituto de Astronom\'\i a, Universidad
Nacional Aut\'{o}noma de M\'{e}xico, Ap. 70-264, 04510 M\'exico, D.F.,
M\'exico.} \altaffiltext{3}{Instituto de Ciencias Nucleares,
Universidad Nacional Aut\'{o}noma de M\'{e}xico, Ap. 70-543, 04510 M\'exico,
D. F., M\'{e}xico.}


\begin{abstract}
It is generally believed that O~stars, confined near the galactic
midplane, are somehow able to photoionize a significant fraction of
what is termed the ``diffuse ionized gas'' (DIG) of spiral galaxies,
which can extend up to 1--\,2\,kpc above the galactic midplane. The
heating of the DIG remains poorly understood, however, as simple
photoionization models do not reproduce the observed line ratio
correlations well or the DIG temperature. We present turbulent
mixing layer models in which warm photoionized condensations are
immersed in a hot supersonic wind. Turbulent dissipation and mixing
generate an intermediate region where the gas is accelerated, heated
and mixed. The emission spectrum of such layers are compared with
observations of Rand (ApJ 462, 712) of the DIG in the edge-on spiral
\ngc. We generate two sequence of models that fit the line ratio
correlations between \sii/\ha, \oi/\ha, \nii/\sii\ and \oiii/\hb\
reasonably well. In one sequence of models the hot wind velocity
increases while in the other the ionization parameter and layer
opacity increases. Despite the success of the mixing layer models,
the overall efficiency in reprocessing the stellar UV is much too
low, much less than 1\%, which compels us to reject the TML model in
its present form.
\end{abstract}





\section{Introduction}\label{sec:intr}

The vast majority of the free electrons in the ISM of the Milky Way
reside in a thick ($\sim 900$\,pc scale height) diffuse layer, which
fills about 20\% of the ISM volume, with a local midplane density of
about 0.1\,\cmc. Such a phase is now known to be a general feature
of external star-forming galaxies, both spirals (e.g., Walterbos
1998; Otte et\,al. 2001, 2002) and irregulars (e.g., Hunter \&
Gallagher 1990; Martin 1997). It is referred to as the warm ionized
medium (WIM) or the diffuse ionized gas (DIG). For edge-on spirals,
only in the more actively star-forming galaxies does the gas
manifest itself as a smooth, widespread layer of emission detectable
above the \hii\ region layer (Rand 1996).
It is generally believed that O\,stars, confined primarily to widely
separated stellar associations near the Galactic midplane, are
somehow able to photoionize a significant fraction of the ISM not
only in the disk but also within the halo, 1--\,2\,kpc above the
midplane. The heating of the WIM or DIG remains poorly understood,
however.  As shown by Madsen, Reynolds \& Haffner (2006), the
temperatures of the Galactic WIM is 2\,000\,--\,3\,000\,K higher
than that of Galactic \hii\ regions. Interestingly, the ratios of
forbidden lines such as \sii, \nii\ or \oi\ with respect to \ha\
have been shown to anticorrelate with the WIM/DIG \ha\ surface
brightness (e.g.  Rand 1998, hereafter R98; Madsen, Reynolds \&
Haffner 2006). The fainter the emission brightness, the higher these
ratios become. A similar trend is found with scale height $z$ above
the disk in edge-on galaxies. Reynolds et\,al. (1999: RHT) propose
that an increase in gas temperature may suffice to explain such
increase in line ratios. The detailed study by Collins \& Rand
(2001) of the DIG in four edge-on galaxies suggests, however,  that
a systematic increase in ionization, at least of the fraction of
O$^++$, might be required as well to account for the observed
trends.

Turbulent mixing layers (hereafter TML) induced by a hot wind is one
of the mechanisms considered by R98 to provide supplemental
ionization and heating. It allowed the author to satisfactorily
reproduce the line ratios observed in the DIG of the edge-on spiral
\ngc\ (situated at $9.5$\,Mpc). In order to obtain a reasonable fit,
however, the mixing layer model had to be combined with a varying
proportion of matter-bounded photoionized condensations. In this
composite model, the photoionization models were taken from J.
Sokolowski (more information can be found in Bland-Hawthorn et\,al.
2001) while the mixing layer calculations were borrowed\footnote{In
particular, the model of SSB with \vt=25\,\kms, $\barT=10^{5.3}$\,K
and depleted gas metallicities.} from Slavin, Shull \& Begelman
(1993: hereafter SSB). In the current work, we propose two
improvements to the heuristic \emph{composite} model introduced by
R98: a) an integrated model: the mixing layer and the matter-bounded
photoionized condensations are calculated as a single gas component
submitted to an external ionizing flux, b) better turbulence
microphysics: rather than considering a single TML temperature given
by the geometrical mean of the warm and hot phases, we derive a
temperature structure for the TML using the mixing length scheme
presented by Cant\'o \& Raga (1991), later implemented in the
emission line code \map\ (Binette et\,al. 1999).

The paper is organized as follows. The equations describing a
turbulent mixing layer and their implementation in the code \map\
are discussed in \S\,\ref{sec:eqn}. The results  obtained from
calculations of the TML structure and of its emission line spectrum
are presented in \S\,\ref{sec:mod} and compared with the line ratios
from the DIG of \ngc. A brief discussion follows in
\S\,\ref{sec:dis}.

\section{Modeling turbulent mixing layers} \label{sec:eqn}

\subsection{Previous works on TML}\label{sec:his}

It is not so clear what exactly the difference is between
``standard'' shock wave heating (i.e., heating and/or ionization due
to the passage of a single, well defined shock wave), and the
heating that takes place in a compressible, turbulent mixing layer.
In particular, it is not entirely clear whether such a turbulent
layer actually generates a number of weak shock waves, or whether it
has a turbulent cascade ending in turbulent dissipation without the
appearance of shock waves.

The initial work on the theory of astrophysical turbulent mixing
layers was presented by Kahn (1980) in the context of Herbig-Haro
jets. Kahn studied the linear and quadratic perturbation theory of
the interface at the edge of a jet flow. The linear theory was
studied by a number of authors in considerable detail in the context
of extragalactic jets in both the gas dynamic and the
magnetohydrodynamic contexts (see, e.g., the review of Bodo 1998).
For the radiative Herbig-Haro jet case, numerical simulations have
been carried out, e.g., by Rossi et~al. (1997) and by Stone, Xu \&
Hardee (1997). Analytic models based on the standard ``turbulent
viscosity'' approach (with a turbulent viscosity parameterized with
a simple, ``mixing length'' approximation) were computed by Cant\'o
\& Raga (1991) and Noriega-Crespo et~al. (1996). Also, Dyson et~al.
(1995), Lizano \& Giovanardi (1995) and Taylor \& Raga (1995)
computed models that included more or less detailed treatments of
the chemistry associated with mixing layers. However, these models
did not include a proper calculation of the ionization state and
line emissivity of the ionized regions of these flows. Binette
et\,al. (1999) adopted the mixing layer scheme developed by Raga \&
Cant\'o (1997) and incorporated the ionization evolution of all the
species considered and the integration of the full emission line
spectrum from the mixing layer. These initial calculations were
intended to explain the spectrum from small excitation knots
observed in Herbig-Haro objects.

In the context of of the interstellar and the intracluster medium,
Begelman and Fabian (1990) proposed a prescription for evaluating
the temperature of a turbulent mixing layer. In a subsequent work,
SSB used the same prescription to calculate the emission line
spectrum from such layers. Their model included radiative transfer,
which allowed these authors to consider the effect of the ionizing
photons emitted by the mixing layer itself. These calculations were
used by R98 in his study of the \ngc\ DIG.

For the current work, we adopted the mixing length prescription of
Binette et\,al. (1999), described below. We also incorporated
radiative transfer, which will allow us to considers the impact of
ionizing photons that originate not only from the internal diffuse
field produced by the mixing layer, but from external sources as
well, such as the ionizing UV from \hii\ regions.

\subsection{A mixing length approach}

\begin{figure}
\resizebox{\hsize}{!}{\includegraphics{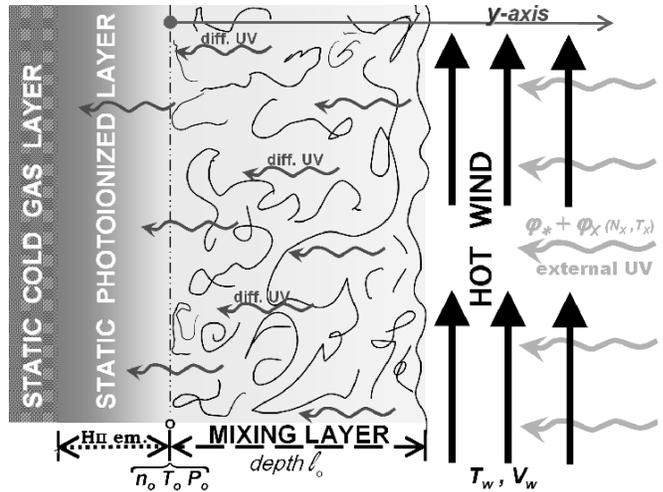}} \caption{Schematic
diagram showing the cross-section along the y-axis of a
plane-parallel mixing layer of thickness \lo. The x-axis (parallel
to the flow) is structure-less in this representation. The mixing
layer is formed by the interaction of a supersonic hot wind (of
temperature \Tw\ and velocity \vw) with a static layer of warm gas
(of H number density \no\ and temperature \To). The gas is entrained
into the mixing layer and possess a velocity $v$, which increases
monotonically from $v=0$ (at $y=0$) up to \vw. The different layers
are all isobaric, with pressure \Po\ ($\approx 2 k \, \no  \To$).
The static layer may contain a photoionized region labeled "\hii\
em." in the figure. Two sources of ionizing photons can be
considered: external UV sources of flux \phis\ due to hot stars and
the X-ray diffuse field generated by the hot wind \phix\ (of column
\Nx\ and average temperature \Tx). The relative importance of the
turbulent and static layers is a free parameter.
} \label{fig:dia}
\end{figure}

For simplicity, we assume an infinite plane interface, along which a
hot wind of temperature \Tw\ is flowing supersonically with velocity
\vw\ with respect to a static warm gas layer of hydrogen number
density \no\ and temperature \To. The turbulent mixing layer that
develops between the hot and warm phases has a geometrical thickness
\lo, as described in the diagram of Fig.\,\ref{fig:dia}.
For the case of a thin, steady state, high Mach number radiative
mixing layer, the advective terms along the direction of the mean
flow can be neglected with respect to the corresponding terms across
the thickness of the mixing layer (see Binette et\,al. 2009 for a
more detailed description). Under this approximation, the momentum
and energy equations can be written as:
\begin{eqnarray}
\mu {d^2 v\over dy^2}=0 \label{eq:vel} \; ,\\
\kappa {d^2T\over dy^2}+\mu \left({dv\over dy}\right)^2=L-G \; ,
\label{eq:flo}
\end{eqnarray}
where $y$ is a coordinate measured from the onset of the mixing
layer (see Fig.~\ref{fig:dia}), $v$ the bulk flow velocity along the
x-axis as a function of $y$, $L$ and $G$ are the radiative energy
loss and gain per unit volume, respectively, and $\mu$ and $\kappa$
are the turbulent viscosity and conductivity, respectively, which
are assumed to be constant throughout the cross-section of the
mixing layer. Eq.\,\ref{eq:vel} can be integrated to obtain the
linear Couette flow solution:
\begin{eqnarray}
v(y)={y\over \lo} \vw \; , \label{eq:vy}
\end{eqnarray}
where \lo\ is the thickness of the TML and \vw\  the mean velocity
of the hot wind (directed parallel to the static warm phase's
surface). This solution can be substituted in Eq.\,\ref{eq:flo},
which can then be integrated to derive the temperature cross-section
across the layer.


Within the TML, we consider the equations governing the
\emph{fractional} abundance $f_i$ of each species $i$. This
abundance must satisfy the equation:
\begin{eqnarray}
D{d^2f_i\over dy^2}=S_f^i \; , \label{eq:ion}
\end{eqnarray}
where $S_f^i$ is the net sink term (including collisional
ionization, radiative and dielectronic recombination, charge
transfer... and processes which populate the current species) of the
species $i$.  The turbulent diffusivity, $D$, is of order unity and
assumed to be position-independent. At the inner and outer
boundaries of the mixing layer, the ionization fractions are  set by
the equilibrium values.

To complete the description of the mixing layer, we require lateral
pressure equilibrium (which determines the density of the flow along
the axis $y$), and calculate the turbulent viscosity with a simple,
mixing length parametrization of the form:
\begin{eqnarray}
\mu=\alpha\,{\rhol}\,\cl\,\lo \; , \label{eq:mu}
\end{eqnarray}
where ${\rhol}$ and \cl\ are the mass density (\gcc) and sound speed
(respectively) averaged\footnote{The subindex $l$ denotes
\emph{averages} across the mixing layer, see Appendix\,\ref{app:a}.}
over the cross-section of the mixing layer, \lo\ is the geometrical
thickness of the mixing layer (Fig.~\ref{eq:vel}).

From the work of  Cant\'o \& Raga (1991), the value of the
proportionality constant $\alpha$ is $0.00247$, as clarified in
Appendix\,\ref{app:a}. This is the required value for a supersonic
mixing layer model to match the opening angle of $\approx 11^\circ$
of subsonic, high Reynolds number laboratory mixing layers in the
limit in which the jet Mach number tends to one. Cant\'o \& Raga
(1991) also find that mixing layer models with this value for
$\alpha$ reproduce the opening angle of mixing layers in jets with
Mach numbers\footnote{It is customary to express \vw\ in terms of
the wind Mach number, $\Mw=\vw/\co$, where \co\ is the adiabatic
sound speed in the static layer.}, \Mw, of up to 20. Our  treatment
of the TML can be extended to cover the subsonic case by replacing
\cl\ in Eq.\,\ref{eq:mu} by \vw, whenever $\vw<\cl$ (see
Appendix\,\ref{app:a}).

Considering that the turbulent conduction and diffusion Prandtl
numbers are of order one, we can compute the conduction coefficient
as $\kappa\approx \mu c_p$ (where $c_p$ is the heat capacity per
unit mass averaged across the mixing layer cross-section) and the
diffusion coefficient as $D\approx \mu/\overline{\rho}$. In this
way, we obtain a closed set of second order differential equations,
Eqs. (\ref{eq:flo}) and (\ref{eq:ion}), which can be integrated with
a simple, successive overrelaxation numerical scheme.

\subsection{TML calculations with the multipurpose code \map} \label{sec:map}

We use the code \map\ (Ferruit et\,al. 1997) to compute the
radiative energy loss term $L$ and the  photoheating term $G$
(Eq.\,\ref{eq:flo}) at each position across the TML. At both the
inner and outer boundaries of the mixing layer, we assume
equilibrium ionization of the different species, while across the
layer, our simple overrelaxation scheme is used to determine the
ionization fractions (Eq.\,\ref{eq:ion}). For the ion diffusion of
each species $f_i$, the spatial differential equations are converted
to temporal equations, with the use of pseudo-time steps $\Delta t =
\Delta{y}^2 ~\lo/(\alpha {\bar n} \cl)$, where ${\bar n}$ is the
average H number density. This allows us to use the temporal
algorithm previously described in Binette \& Robinson (1987) for
determining the spatial diffusion of the ionic species.

The radiative transfer is determined by integrating (from the hot
layer $y=\lo$ down to $y=0$) the intensity of the UV diffuse field
that the layer produces, assuming the {\it outward only}
approximation. Any external UV radiation impinging from the external
side is simply added to the diffuse field at the onset of the
integration. The intensity of the external ionizing field is defined
by the ionization parameter as follows:
\begin{eqnarray}
\Uo= \phiw/c\,\no \; , \label{eq:uo}
\end{eqnarray}
where $c$ is the speed of light, \phiw\ the flux number of ionizing
photons impinging on the TML, and \no, the total H density at $y=0$.
The flux \phiw\ is the sum of two components: \phis\ resulting from
external UV sources such as hot stars and \phix\ due to the X-ray
diffuse field generated by the hot wind (of column \Nx\ and average
temperature \Tx). The absorption processes that are taken into
account in the transfer equations across the TML include all the
relevant photoelectric cross-sections of the ions present.

Collins \& Rand (2001) used a method based on \niiw\ to estimate the
gas temperature and, by bootstrapping, they could infer the
ionization fractions of N and other species. They find that the
hydrogen ionization fraction in \ngc\ is as high\footnote{Comparable
values were inferred for the DIG in the Milky\,Way by Reynolds
et\,al. (1998) using \oiw.}  as 0.55--\,0.80. Because H is fairly
ionized and since X-ray ionization is very inefficient in comparison
with UV radiation, we infer that their role relative to the soft UV
flux from stars must be relatively unimportant. We therefore
neglected the term \phix\ by setting \Nx\ to a negligible value.

For all the calculations presented in this work, the mixing layers
are considered immersed in the radiation field that escapes from
\hii\ regions located in the disk. This picture assumes that disk
\hii\ regions are on the whole matter-bounded. According to Beckman
et\,al. (2000), Zurita et\,al. (2002) and Rela\~no et\,al. (2002),
the fraction of Lyman photons that escape giant \hii\ regions
(labeled ``leakage fraction'' in \S\,\ref{sec:sed}) may lie in the
range of 30--\,50\%. The impact of photoionization on our TML model
can be inferred from the internal behavior of \gam\ across the
layer, where \gam\ is defined as follows:
\begin{eqnarray}
\gam={{L-G}\over{L+G}} \; . \label{eq:gam}
\end{eqnarray}
The quantity \gam\ is zero when the temperature corresponds to the
equilibrium value and tends toward unity when cooling becomes much
stronger than heating. In our calculations, we iteratively determine
\To\ at $y=0$ until $\gam=0$ is obtained.

After computing the emission line spectrum of a given TML, \map\
offers the option of separately computing the emission lines
generated by the inner photoionized layer (i.e. $y<0$ in
Fig.\,\ref{fig:dia}) where equilibrium ionization prevails. A simple
isobaric photoionization model is calculated in this case, using the
radiation field that has \emph{not} been absorbed by the mixing
layer. The total line spectrum from all the layers is then given by
the \emph{sum} of the line intensities from both the TML and  the
static equilibrium photoionization calculation.
For all the models studied, however, whenever a static photoionized
layer was included, it dominated the total spectrum and dwarfed the
spectral signature of the turbulent layers. In order to enhance the
contribution of the TML and successfully fit the ratios observed in
\ngc, the static equilibrium region must be obviated and the total
thickness of the emission layer becomes equal to that given by the
turbulent layer alone. This might come about in a 3D geometry if the
cold gas were to reside in the core of dense condensations whose
\emph{area} is much smaller than that of the surrounding turbulent
layers.


Since the mixing layer is isobaric, with pressure \Po, the density
profile depends on the behavior of the temperature within the layer
itself. The value of \Po\ is set by two quantities evaluated at the
inner TML boundary ($y=0$): the density \no\  and the equilibrium
temperature \To. To be definite, a density of $\no=1$ was adopted
that applies to all models presented in this work. At such a low
density, the {\it low density regime} fully applies and the
calculated line ratios are therefore independent of the actual \no\
value used, provided the product \no\lo\ is conserved. Our models
are appropriate for any density $\no\la 50$. When the external
parameters (\Uo, \Tw\ and \vw) are kept constant, TML models turn
out to be equivalent whenever the product of the H\,density \no\ and
thickness \lo\ is the same. Therefore, at least in the low density
limit regime typical of the DIG, it is sufficient to specify the
quantity\footnote{The quantity \No\ is a convenient model
descriptor. However, it is a bad estimator of the true integrated H
column, which is a lot smaller since the density is not constant but
decreases as the temperature rises with thickness $y$
(Fig.\,\ref{fig:dia}).} \No=\no\lo\ to uniquely define a model.

We found that the temperature of the hot wind is not a critical
parameter of the TML models. Its value can be varied and still
result in a sequence of equivalent models (i.e. with similar line
ratios), provided the value of \vw\ is kept constant. For this
reason, we adopted the same value of $\Tw=8\times 10^{5}\,$K for all
the models presented in this work.

To  summarize, in order to compute solutions to the mixing layer, we
must first define the external radiation field and then specify the
values of the following parameters: the mixing layer's \No, the
ionization parameter, \Uo, and finally the velocity of the hot wind,
\vw, and its temperature, which we kept constant at $8\times
10^{5}\,$K.



\section{Modeling the DIG in \ngc} \label{sec:mod}

\subsection{The gas abundances}\label{sec:met}

Unless otherwise specified, the abundances of the elements are solar
(Anders \& Grevesse 1989) and have the following values: (He, C, N,
O, Ne, Mg, Si, S, Ca, Ar, Fe) $= [10^5, 363, 112, 851, 123, 38, 36,
16, 2.3, 3.6, 47] \times 10^{-6}$ relative to H (by number).
Depleted or oversolar metallicities resulted in line ratios
difficult to reconcile with the forbidden line ratios observed in
\ngc.


\begin{figure}
\resizebox{\hsize}{!}{\includegraphics{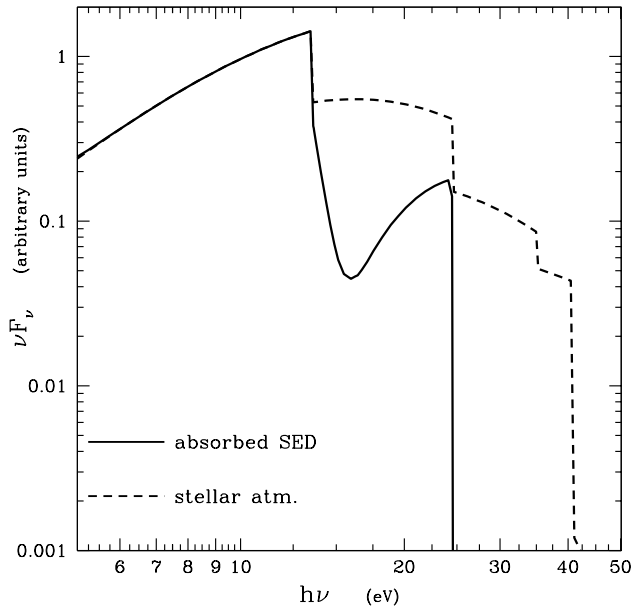}} \caption{The \sed\
of a 38\,000\,K star (dashed line) as a function of photon energy.
The solid line represents the \sed\ of the UV radiation escaping
from a matter-bounded photoionized slab that absorbs 80\% of the
impinging ionizing photons. It is assumed that the DIG is exposed to
this absorbed \sed.} \label{fig:mod}
\end{figure}

\subsection{The ionizing spectral energy
distribution}\label{sec:sed}

The first step consists of defining the spectral energy distribution
(\sed) of the ionizing radiation field \emph{reaching} the DIG. A
useful constraint is provided by the observed \heiw/\ha\ ratio. In a
previous observational work on \ngc, Rand (1997) measured a ratio of
$\heiw/\ha \approx 0.034$, which he estimated implies a ratio of He
ionizing photons to H ionizing photons, \QH/\QHe, of about 0.08. To
be consistent with this datum,  Rand favors a stellar temperature of
37\,500\,K. In the calculations presented here, we adopt a similar
temperature of 38\,000\,K. The atmosphere models used in \map\ are
from Hummer \& Mihalas (1970). As in R98, we assume that only a
fraction of the stellar radiation escapes the \hii\ regions in the
spiral disk to reach the DIG. To be definite, we have set the
leakage fraction to 20\%. To compute this modified \sed, we have
computed a plane-parallel photoionization model and extracted the
photon spectrum at a depth where 80\% of the ionizing photons have
been used up. The diffuse ionizing radiation from the slab itself
has been included. The resulting \sed\ is shown in
Fig.\,\ref{fig:mod} (solid line). Because this modified \sed\
contains few ionizing photons beyond 24.6\,eV, the mixing layer
model presented below is characterized by an $\heiw/\ha$ ratio as
low as $\simeq 0.003$. This suggests a higher stellar temperature
for \ngc\ than considered here and in Rand (1997), or a larger
leakage fraction than assumed above.

We have verified that the stellar temperature can be raised up to
40\,000\,K without exceeding the observed \hei/\ha\ value (assuming
the same leakage fraction). For the DIG in the Milky Way, \heiw/\ha\
is observed to be significantly smaller, lying in the range
0.01--\,0.04 (Madsen et\,al. 2006).



\begin{figure}
\resizebox{\hsize}{!}{\includegraphics{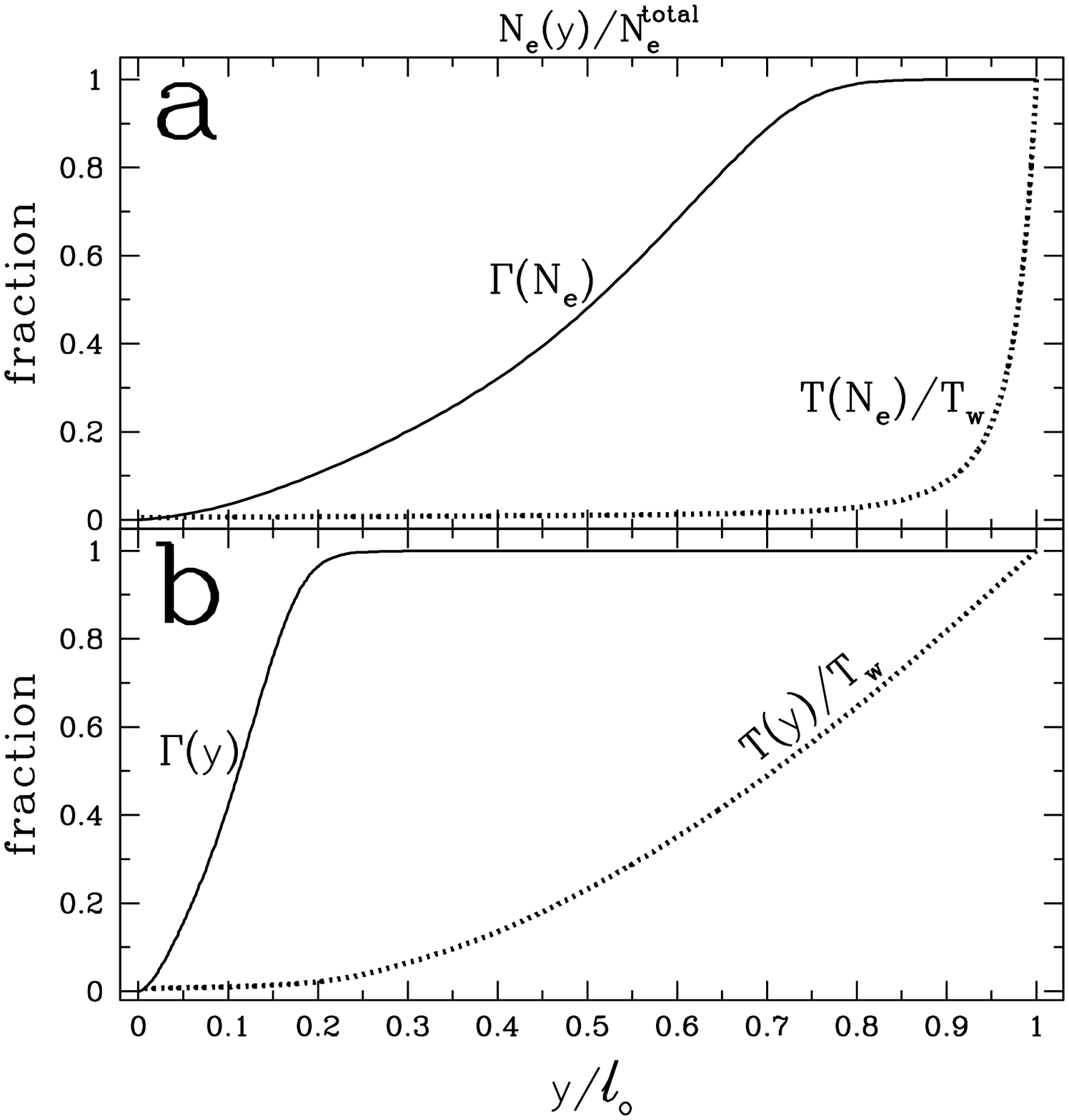}} \caption{Behavior of
\gam\ (Eq.\,\ref{eq:gam}) and of the local temperature as a function
of (\emph{a}) the electron column density (normalized to the total
value of $1.1\times 10^{16}$\,\cms) and of (\emph{b}) the normalized
thickness $y/\lo$. The model shown has \vw=85\,\kms, $\Tw=8\times
10^{5}$\,K, $\Uo=10^{-5}$ and $\No=10^{17}$\cms. The temperature
scale is normalized relative to \Tw.
The value of the equilibrium temperature, at the onset of the layer,
is \To=4390\,K (i.e. $\To/\Tw=0.0055$).
} \label{fig:gam}
\end{figure}


\begin{figure}
\resizebox{\hsize}{!}{\includegraphics{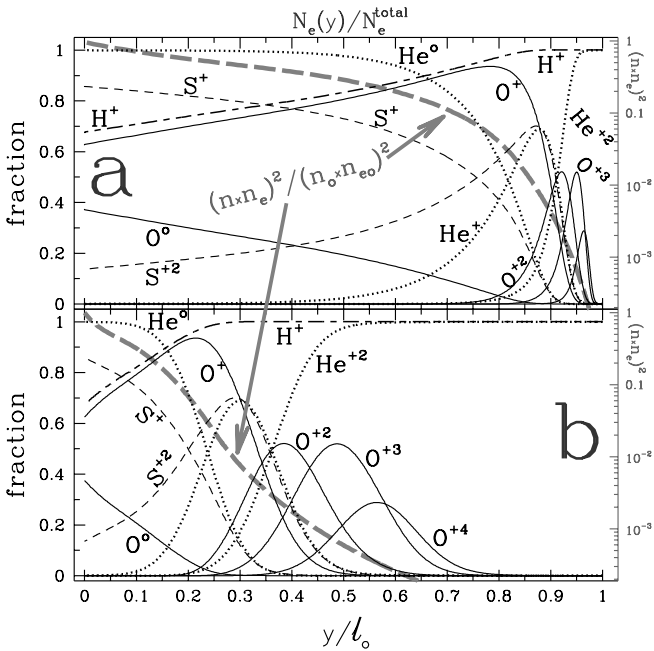}} \caption{Ionization
fractions of O, S, He and H  for the same model as in
Fig.\,\ref{fig:gam} as a function of (\emph{a}) the electron column
density (normalized to the total value of $1.1\times 10^{16}$\,\cms)
and of (\emph{b}) the normalized thickness $y/\lo$. Also shown is
the quantity $(n\times n_e)^2$ (thick grey dashed line), which is
normalized to the value at $y=0$ and is plotted on a logarithmic
scale, to be read using the rightmost axis. The volume line
emissivities from each ion is proportional to this factor. The
lines'\,coding is the same in both panels.
} \label{fig:ion}
\end{figure}

\subsection{TML temperature and ionization structure}\label{sec:stru}

In Fig.\,\ref{fig:gam} we  present the behavior of the temperature
as well as of the imbalance between cooling and heating (\gam, see
Eq.\,\ref{eq:gam}) as a function of thickness $y$ within a
representative mixing layer calculation. For this model, the hot
wind velocity is 84\,\kms, while the ionization parameter is
$\Uo=10^{-5}$ and the layer's column $\No=10^{17}$\cms. The inner
boundary is characterized by an equilibrium temperature of
$\To=4\,390$\,K. As illustrated in Fig.\,\ref{fig:gam}, due to the
very hot wind temperature ($\Tw \gg \To$),  \gam\ increases
monotonically to reach unity at $y=\lo$ as a result of the cooling
term becoming progressively dominant relative to the photoheating
term.

The ionization structure within the same TML is shown in
Fig.\,\ref{fig:ion}. The static layer if it was present would have
ionization fractions quite similar to those found at the extreme
left in this figure. Line intensities depend on temperature, but
scales also with density square. Since the quantity $(n\times
n_e)^2$ decreases strongly with increasing thickness (see thick grey
dashed line), the local emissivities likewise decrease markedly
towards the right. It is therefore apparent that any emission from
\oiii\ or any other species with similar or higher ionization
potential are produced in this model only within the mixing layer,
and not within the static layer. This would of course be different
if a significantly higher stellar temperature was chosen.


\subsection{DIG line ratio trends in \ngc}\label{sec:rat}

For comparison with TML models we selected the work of R98 on the
DIG in \ngc. The data on this object present the following
advantages: the spectral data are homogeneous, they are of
sufficient quality to reveal subtle intrinsic correlations among the
line ratios observed at different positions across and along the
disk of this edge-on spiral, finally, mid-infrared halo and disk
observations with the Spitzer Space Telescope are now available
(Rand et\,al. 2008). We postulate that the line ratio correlations
are a direct manifestation  of the DIG excitation mechanism rather
than positional changes in gas metallicity or in effective stellar
temperature.

To optimize the comparison with models, we present the line ratios
in Fig.\,\ref{fig:seqa} in a format which reveals the essential
trends found by R98 in his long slit data. The three ratios
\oiii/\hb\ (5007/4861), \oi/\ha\ (6300/6563) and \nii/\sii\
(6583/6716) are all plotted against the same ratio \sii/\ha\
(6716/6563). The \oiii/\hb\ line ratio lies below unity in all
measurements (panel\,{\it a}). What is most striking about the data
is that not only do the forbidden line intensities \oi, \nii\
increase along with \sii/\ha, but the high excitation \oiii/\hb\
ratio as well. Such a behavior runs counter to the predictions of
simple photoionization models where \oiii/\hb\ increases markedly
when the ionization parameter is increased, while the ratios
\sii/\ha, \oi/\ha\ or \nii/\ha\ all decrease. This leads to the
proposition by R98 and Collins \& Rand (2001) that another
ionization process besides photoionization is likely to be involved.
With the Milky Way DIGs, the derived temperatures are found to
increase along with the increase of the forbidden lines relative to
\ha\ or \hb\ (e.g. Madsen et\,al. 2006). According to RHT, a
systematic temperature increase might be all that is necessary to
reproduce the increase in the forbidden line intensities. Again,
this runs counter to predictions of simple photoionization models
where the temperature is quite insensitive to changes in the value
of \Uo. Hence various authors have concluded in the existence of a
new heating (e.g. RHT) and possibly ionizing mechanism (e.g. R98) in
addition to photoionization. For the sake of completeness, we
included an inset in Fig.\,\ref{fig:seqa} (panel\,{\it d}) where the
\nii\ temperature computed in models is plotted against \sii/\ha. We
may reasonably expect that any realistic DIG models should show a
significant temperature gradient in such a diagram.

\subsection{Specific TML calculations}\label{sec:cal}

\begin{figure}
\resizebox{\hsize}{!}{\includegraphics{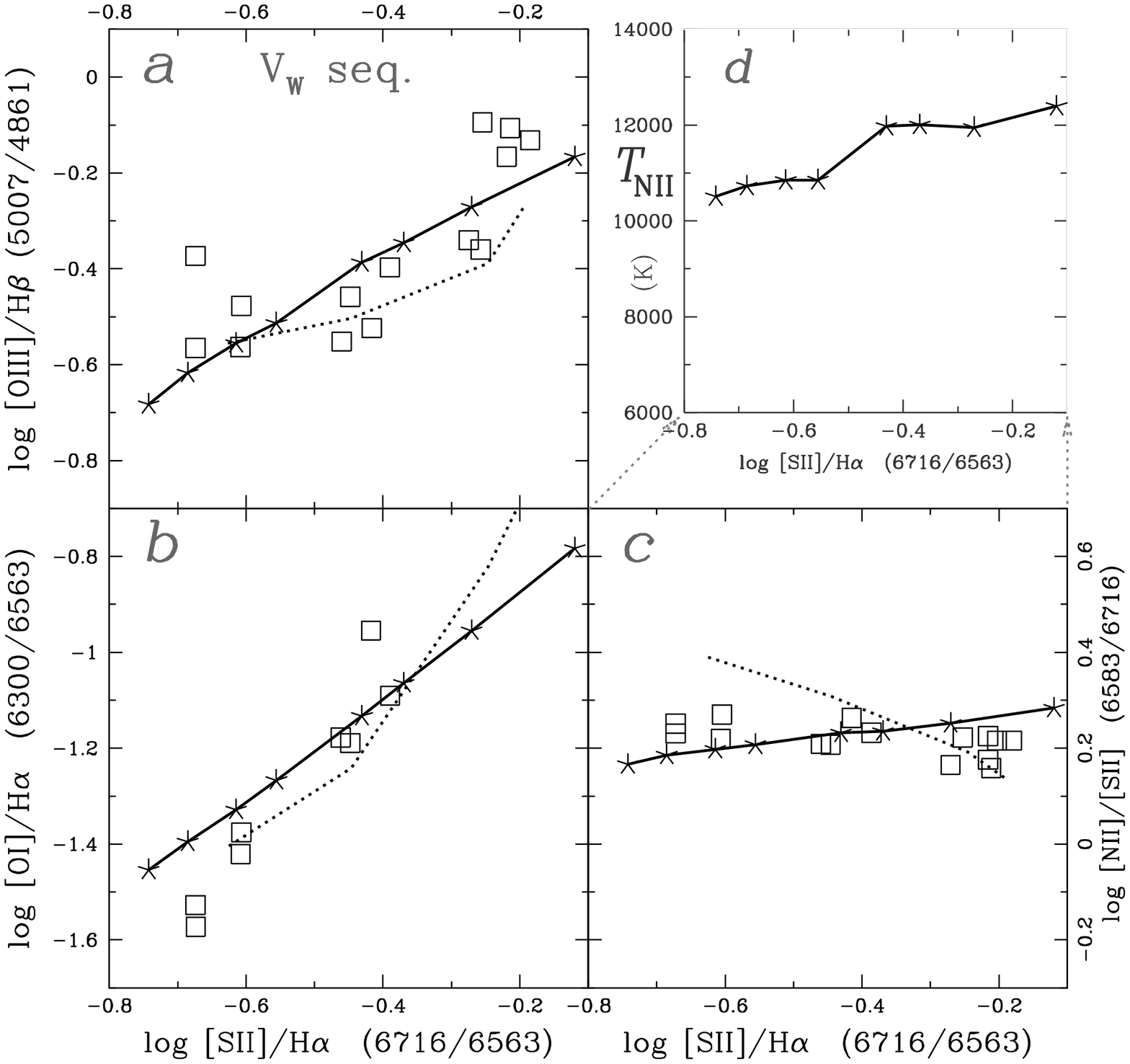}} \caption{Line ratios
of (\emph{a}) \oiii/\hb, (\emph{b}) \oi/\ha, (\emph{c}) \nii/\sii\
and (\emph{d}) the \Tnii\ temperature, all as a function of the
\sii/\ha\ line ratio. Open squares: measurements of R98 of the DIG
in \ngc; continuous line: a sequence of TML models in which \vw\ is
progressively increased by 0.1\,dex from 27 to 133\,\kms, while
other parameters are kept constant: $\No=10^{17}$\,\cms\ and
$\Uo=10^{-5}$. The ratio \sii/\ha\ \emph{increases} with increasing
\vw. The dotted line represents a varying mixture of turbulent
mixing layers and photoionized matter-bounded models that were
proposed by R98. } \label{fig:seqa}
\end{figure}

We carried out an extensive exploration of the parameter space in
order to determine the parameter range that could best reproduce the
trends present in Fig.\,\ref{fig:seqa}. As in previous studies, we
concur that the layer has to be matter-bounded, otherwise the
\oiw/\siir\ ratio would significantly exceed the observed values.
Furthermore, for the calculations to show a significant departure
from steady-state photoionization as a result of turbulent
dissipation and ionization mixing, it required an extremely thin
layer, with a column \No\ of order $\sim 10^{17}$\,\cms\ only. This
raises concerns about the efficiency of turbulent dissipation to
contribute significantly to the DIG emission, as discussed in
\S\,\ref{sec:dis}. Because the TML layer in the composite model of
R98 was decoupled from the photoionized component, the author was
free to  adopt a much larger column of $2\times 10^{18}\,$\cms\ in
his matter-bounded photoionization calculations.


\begin{figure}
\resizebox{\hsize}{!}{\includegraphics{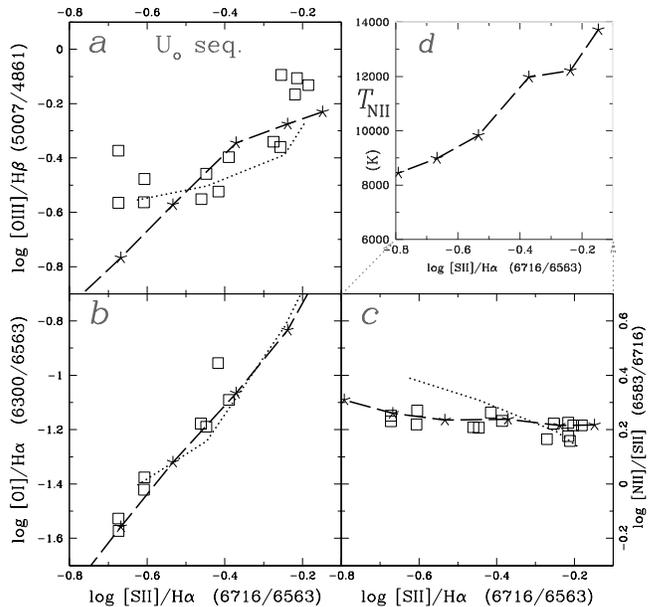}} \caption{Dashed
line: a sequence of TML models in which the ionization parameter is
progressively increased by 0.2\,dex, from $4.0\times 10^{-6}$ to
$4.0\times 10^{-5}$. The column \No\ increases in the same
proportion, from $0.4\times 10^{17}$ to $4.0\times 10^{17}$\,\cms\
while \vw\ is kept constant at 85\,\kms. The ratio \sii/\ha\
\emph{decreases} with increasing \Uo\ and \No. Other symbols have
the same meaning as in Fig.\,\ref{fig:seqa}.} \label{fig:seqb}
\end{figure}

A characteristic of the DIG is its relatively weak \oiii\ line
intensity, which favors relatively small values for the ionization
parameter when considering photoionization (e.g. Sokolowski \&
Bland-Hawthorn 1991). In the case of the composite TML models of
R98, the author has employed values that varied between $10^{-5}$ to
$8\times 10^{-5}$. In the TML regime that we have explored, we are
lead towards similar values, with \Uo\ typically at $\sim 10^{-5}$.
We find that the \oiii\ flux is the result of turbulent dissipation
rather than of photoionization. On the other hand, low ionization
species responsible for lines such as \oii, \nii, \sii... are mostly
caused by photoionization, but the additional heating due to
turbulent dissipation causes these lines to become much brighter. In
fact, if a static layer (see Fig.\,\ref{fig:dia}) with such a low
value of \Uo\ was present behind the TML, the overall line ratios
would lie near \nii/\sii=0.93, \sii/\ha=0.087 and \oi/\ha=0.013.
Therefore, these ratios would all lie outside the boundaries from
the three line diagrams \emph{a}, \emph{b} and \emph{c} of
Fig.\,\ref{fig:seqa}. If we relied only only photoionization
(without any TML) to fit the observations, we would have to adopt a
much harder ionizing \sed\ (but see \S\,\ref{sec:sed}) to provide
the necessary heating and the ionization up to O$^{+2}$.

\subsubsection{The \vw--\,sequence}

In Fig.\,\ref{fig:seqa}, we superpose a sequence of TML models to
the data, in which the wind velocity is progressively increased in
locked steps, by 0.1\,dex, from $\vw=27$ to 133\,\kms\
(corresponding to Mach numbers increasing from \Mw=2.53 to 13.5).
The column \No\ was set to $10^{17}$\,\cms\ and \Uo\ to $10^{-5}$.
In that sequence, the forbidden line ratios of \oiii/\hb\ and
\sii/\hb\ increase with increasing \vw. We note that the temperature
inferred from the \nii\ (5755/6583) line ratio, \Tnii, increases
slightly along with \sii/\ha, as shown in the upper right inset
(panel $d$). Overall, the trends in line ratios are well reproduced,
better in fact than with the composite (photoionization+TML) model
of R98, which is represented by the dotted line. R98 showed that the
line ratios not only correlate with decreasing \ha\ surface
brightness, but also with increasing $z$ height above the disk of
\ngc. If the \vw\ sequence of models were a valid description of the
long-slit measurements of R98, this would imply that the wind
velocity is increasing with $z$.

\subsubsection{The \Uo--\,sequence}

Alternatively, it is  possible to reproduce the line ratio
correlations by varying the ionization parameter instead of \vw. In
Fig.\,\ref{fig:seqb}, we show a sequence of models in which both
\Uo\ and \No\ increase\footnote{Such a sequence is achieved by
increasing both  quantities \phiw\ and \lo\ in lock steps, while
keeping \no\ constant.} in lock steps by 0.2\,dex, while \vw\ is
kept constant at 85\,\kms. The initial values of \Uo\ and \No\ are
$4.0\times 10^{-6}$ and $0.4\times 10^{17}$\,\cms, respectively,
while the final values are 10 times larger. Varying \No\ in locked
steps with the ionization parameter limits the variations in \hi\
opacity, which is desirable otherwise the TML layers either get too
thin or reach a regime in which photoionization takes over. In the
sequence shown in Fig.\,\ref{fig:seqb}, we have ensured that
photoionization does not become the dominant process. In fact,
\oiii\ and the other forbidden lines actually decrease as \Uo\ is
increased. For instance, \oiii\ gets weaker as \Uo\ increases
because the average gas temperature where it is produced
\emph{decreases} along the sequence from 51\,500 to 40\,300\,K. This
sequence of models shows a \nii\ temperature gradient
(Fig.\,\ref{fig:seqb}{\it d}) that is stronger than in the previous
sequence (Fig.\,\ref{fig:seqa}{\it d}). When compared to the data on
\ngc, the behavior of the \oiii/\hb\ ratio is not as closely fit as
in the \vw--\,sequence, but the fit of the other line ratios are
better.


\begin{figure}
\resizebox{\hsize}{!}{\includegraphics{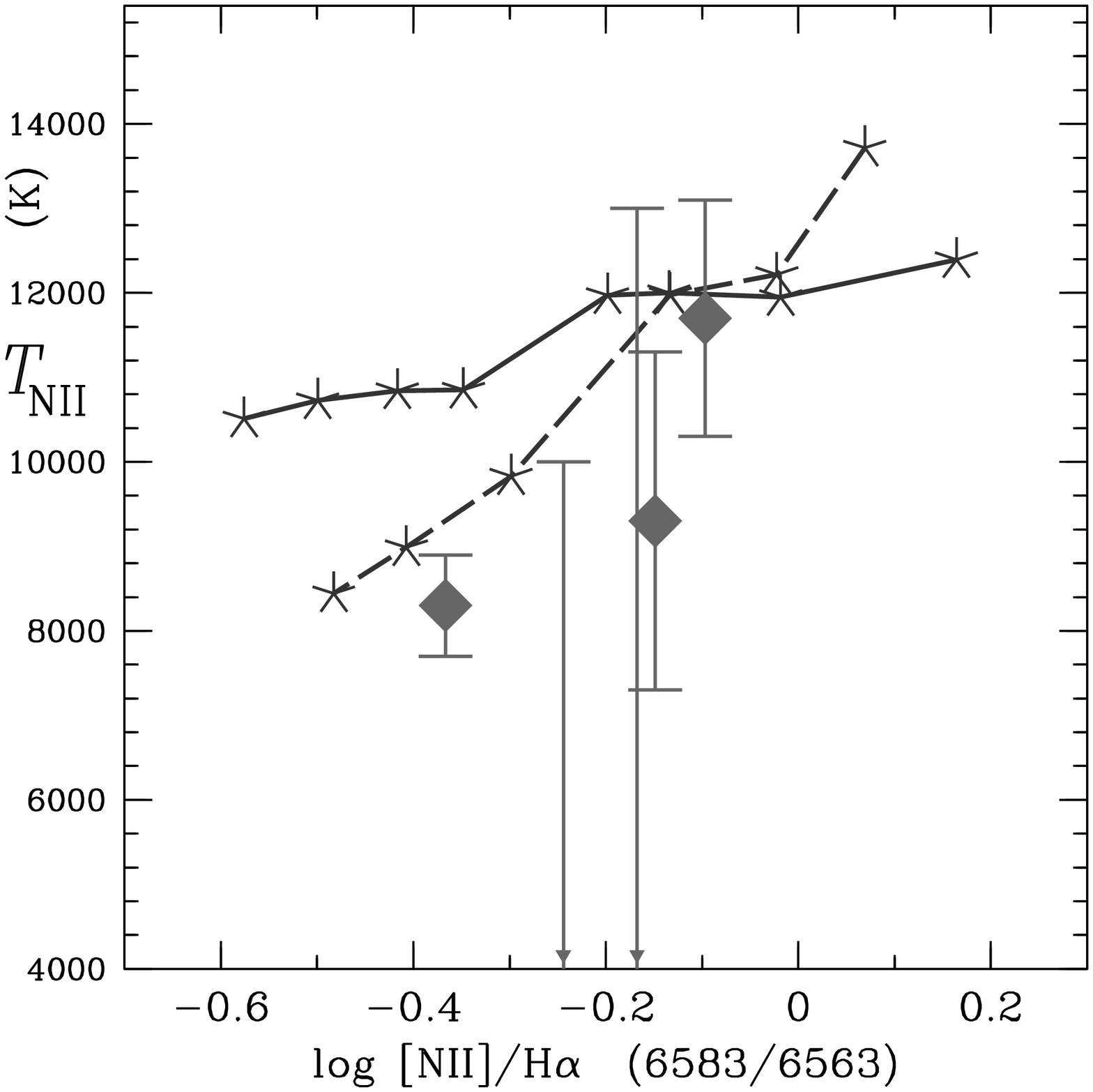}} \caption{The \Tnii\
temperature was derived from the temperature sensitive \nii\
(5755/6583) line ratio as a function of the \nii/\ha\ line ratio.
The three solid squares correspond to temperatures borrowed from of
Madsen et\,al. (2006) and Reynolds et\,al. (2001) of the DIG in the
Milky Way while the two upper limits correspond to \ngc\ (Rand
1997).  Overlayed are the same \Uo\ and \vw\ sequences of
Fig.\,\ref{fig:seqa} (solid line) and Fig.\,\ref{fig:seqb}
(dashed-line), respectively.} \label{fig:tnii}
\end{figure}

\subsection{Other DIG observations}\label{sec:oth}

Although this study focuses on \ngc, fitting similar line ratio
trends in other objects is certainly achievable. In itself, this
does not validate TML models, since the main reason for this is the
higher number of free parameters that can be varied and which were
already numerous in the alternative case of pure photoionization.

Although forbidden line ratios with respect to \ha\ or \hb\ increase
with height $z$ above the midplane in \ngc, this behavior is not
repeated in all spirals. Cases for instance where \oiii/\hb\
decreases or reverse trend with height $z$ are well documented in
the studies of Miller \& Veilleux (2003), Otte et\,al. (2002) and
Collins \& Rand (2001). In their studies of face-on M33, Voges \&
Walterbos (2006) find sudden changes in this ratio. Within the
perspective of TML models, these observations bring new constraints
to the interpretations that models are able to provide. For
instance, in cases where \oiii/\hb\ decreases with \emph{z}, we find
it improbable that \vw\ could be decreasing with height, as required
by the \vw--\,sequence (Fig.\,\ref{fig:seqa}). In the case of the
\Uo--\,sequence (Fig.\,\ref{fig:seqb}), however, a TML
interpretation would imply that \Uo\ is increasing with height,
which could only come about if the ambient gas pressure decreases
faster with \emph{z} than the local ionizing flux.

The direct measurement of the \nii\ temperature is difficult owing
to the intrinsic weakness of the \niitw\ lines. The \nii\
(5755/6583) line ratio has nevertheless been determined in the Milky
Way for three DIG sight lines  (Madsen et\,al. 2006; Reynolds
et\,al. 2001). The inferred temperatures are plotted in
Fig.\,\ref{fig:tnii}.  The two upper limits for \ngc, as determined
by Rand (1997), are also plotted. The comparison of these three
temperatures by Madsen et\,al. (2006) with those of `classical'
\hii\ regions confirms that the DIGs are approximately
2000--\,3000\,K warmer than \hii\ regions. The TML models presented
in this work are overlayed in Fig.\,\ref{fig:tnii}. Despite the
scarce data, the comparison with models is encouraging insofar as it
shows that turbulent dissipation and mixing can provide the extra
heating required by the Galactic DIGs.

With the TML models presented in this work, in which none of the
\oiii\ emission is the result of photoionization but rather due to
turbulent mixing and dissipation, we find that \oiii\ is produced at
a depth $y$  of very high electron temperature. The behavior of the
\oiii\ (4363/5007) line ratio in models shows that in both sequences
the increase in \oiii\ is caused by an increase in
average\footnote{Because of the steep temperature gradient across
the TML, the integrated \oiii\ (4363/5007) line ratio represents an
`average' temperature that is strongly biased towards layers where
the gas is densest rather than being close to the temperature where
the fraction of O$^{+2}$ peaks (see behavior of $(n\times n_e)^2$ in
Fig.\,\ref{fig:ion}).} temperature, from $\simeq 40\,000$ to
52\,000\,K. These values are much higher than those characterizing
the \nii\ emission zone, to the extent that the \oiiitw\ line flux
is between the value of the \niitw\ flux and half of it. In
principle, measuring a high \oiii\ temperature or not would provide
a strong discriminant for or against TML models.

\subsection{The mid-infrared \neiiw\ and \neiiiw\ lines}\label{seq:ir}

Mid-infrared observations of \ngc\ using the Spitzer Space Telescope
has been reported by Rand et\,al. (2008). Of particular interest are
the measurements of the \neiiw, \neiiiw\ and \siiiw\ fluxes of the
DIG in the halo (2\,measurements) as well as in the disk
(1\,measurement). Interestingly, the observed \neiii/\neii\ ratio
increases with height $z$ by a factor of three. This is
qualitatively similar to the trends revealed by the optical ratios
of \sii/\ha\ or \oiii/\hb. Our two TML sequences has \neiii/\neii\
increasing along with \oiii/\hb, but the increase is only a factor
two. The maximum values reached for the ratio \neiii/\neii\ are
0.067 and 0.070, for the \vw--\,sequence and \Uo--\,sequence,
respectively. This is about 4.5 times too weak with respect to the
observed halo value of 0.31. In his analysis, Rand et\,al. (2008)
adopted a Ne/O abundance 2.5 times higher ratio than the solar value
(\S\,\ref{sec:met}). Changing our abundance set  might resolve this
discrepancy.

\section{Discussion}\label{sec:dis}

A photoionized turbulent mixing layer can reproduce the line ratio
trends observed by R98, if it is matter-bounded. Our work confirms
the initial intuition of R98 who mixed two causally unrelated
ionized components in ad\,hoc proportions: a TML component and a
matter-bounded photoionized component. The main advantage of our
integrated model is that the ionized slab considered is submitted in
a self-consistent manner to both excitation mechanisms: turbulent
dissipation and photoionization. Adjusting the parameters so that
the observed line ratios can be reproduced, however, comes at a
price. The thickness of the TML layer becomes so small that the
overall efficiency in generating emission lines becomes a concern.
This can be illustrated as follows. From the definition of \Uo, the
ionizing photon flux impinging on our putative mixing layer of area
$A$ is $3.0\times {10}^{4} \, A\, \nou \, \Uf$\, \qts, where \Uf\ is
$\Uo/10^{-5}$ and $\nou=\no/0.1\,\cmc$. In the case of pure
photoionization, assuming ionization equilibrium, the \ha\
luminosity is given by $\Lha = 4.44\times 10^{-8} \effha\, A\,
\nou\, \Uf$\,\ergs\ where \effha\ is a measure of the reprocessing
efficiency, that reaches unity in the case of an
\emph{ionization-bounded} layer, where all the ionizing photons are
absorbed. When the \Lha\ from the TML models are compared to that
expected from this simple relationship, we find that the
\vw--\,sequence is characterized by an efficiency \effha\ that goes
from 0.013 to 0.0031, as \vw\ is increased. In the case of the
\Uo--\,sequence, \effha\ goes from 0.0013 to 0.027, as \Uo\ and \No\
are increased. What is most problematic, is that the efficiency is
lowest when turbulent dissipation is maximized, that is, in the
regime where the \oiii/\hb\ ratio is highest, whether we consider
the \vw--\,sequence of Figs.\,\ref{fig:seqa} or the \Uo--\,sequence
of Fig.\,\ref{fig:seqb}. The efficiencies at the high \oiii/\hb\ end
are only 0.0031 and 0.0013, for the \vw--\,sequence and
\Uo--\,sequence, respectively.

Such an efficiency is uncomfortably low. This can be illustrated by
using the observations of \ngc\ by Rand (1997). With the long-slit
perpendicular to the galactic plane of edge-on \ngc, the author
measured an \ha\ surface brightness of $2\times 10^{-17}\,$ \flusur\
at an height $z$=2\,kpc, where the \oiii/\hb/ ratio peaks. Assuming
a distance of $D=9.5$\,Mpc (1\,arcsec=46\,pc), the slit,
perpendicular to the galactic plane,  crosses the mid-plane of \ngc\
at 4.6\,kpc from the nucleus. We adopt 1\arcsec\ as the reference
spatial scale. We integrate along the thickness of the layer, which
we assume extends at most over 9\,kpc along the line of sight, so
that the layer area \Ap\ exposed to the UV flux arising from the
midplane becomes 46\,pc\,arcsec$^{-1}$\,$\times$\,9\,000\,pc. The
surface brightness derived from the \Lha\ expression above becomes:
$\Fha = \Lha \, k\, \Ap \, /(4\pi D^2)$, where $k$ is the number of
projected TML layers  per arcsec (at $z$=2\,kpc) along the slit.
Hence, $\Fha= 1.6\times 10^{-17} k\, \effha\, \nou\, \Uf$\,\flusur.
For an efficiency \effha\ of 0.0013 and with \Uf=0.4, matching the
observed surface brightness implies the unreasonable number of
$k\simeq 2400$ layers\footnote{For this particular model, the
layer's geometrical thickness is $\lo \simeq 4\times
10^{17}/\nou$\,cm, which in any case leaves no space for the hot
wind to flow if $k$ is large.} per arcsec along the slit, assuming a
density typical of the DIG in the Milky Way (\nou=1). Even if one
raises this density by an order of magnitude, too many layers are
still required. For higher density values beyond that, there would
not be enough O stars to account for the required ionizing flux.

To summarize, it is possible to contrive the values of the
parameters describing TML models so that  the observed trends in
line ratios are reproduced reasonably well. However, the
efficiencies of any successful models appear downright too
insignificant. In its current form our TML model is therefore
invalidated. We may speculate that the gradient in temperature
between the hot wind and the warm photoionized gas is so steep that
the mixing length approach explored by Cant\'o \& Raga is not
directly applicable to the very hot wind surrounding the DIG. Values
for the constant $\alpha$ (Eq.\,\ref{eq:mu}) about two orders of
magnitude larger would be required so that turbulent mixing and
dissipation could operate over a layer's thickness much larger than
the very small values found in the current model.


\acknowledgements One of us, L.\,D., acknowledges financial support
from the Canada Research Chair program and from Canada's Natural
Science and Engineering Research Council (NSERC) and from Qu\'ebec's
``Fonds qu\'eb\'ecois de la recherche sur la nature et les
technologies'' (FQRNT). This work was supported by the CONACyT grant
J-50296.
Diethild Starkmeth helped us with proofreading.


\appendix

\section{The derivation of constant $\alpha$}\label{app:a}

From Eq.\,A3 in Cant\'o \& Raga (1991), we obtain that the turbulent
viscosity coefficient $\mu$ is given by:
\begin{eqnarray}
\mu=  {\epsilon\over 12} \rho_o c_o \,\lo \; ,\label{eq:mur}
\end{eqnarray}
where  $\epsilon$ is given by $0.089 \times {\rm MIN}[\epsilon_1,
\epsilon_2]$ with $\epsilon_1 = {1 \over 2} {c_o\over c_w}$ and
$\epsilon_2 = {1 \over 3} {c_o\over c_l}$ (see their Eqs.\,27, 30
and 36). $c_w$ and $c_o$ correspond to the adiabatic sound speed in
the hot wind and in the static layer, respectively. The subindex $l$
denotes \emph{averages} across the mixing layer. Cant\'o \& Raga
(1991) refer to the $\epsilon_2<\epsilon_1$ case as the `mixing
layer limited' case while the `jet limited' case corresponds to
$\epsilon_2>\epsilon_1$

In the case where $\epsilon_2 < \epsilon_1$, since in the isobaric
case $\rhol c_l^2 = \rho_o c_o^2$, after multiplying the numerator
and denominator by $\rhol c_l$, we can rewrite the above expression
as
\begin{eqnarray}
\mu= {0.089 \over 3 \times 12} \rhol c_l \lo \; .\label{eq:mue}
\end{eqnarray}
In the notation used in this Paper, the constant $\alpha$ simply
corresponds to the ratio $0.089/36=0.00247$. In the transonic case
or when $\epsilon_1<\epsilon_2$, we replace $c_l$ in
Eq.\,\ref{eq:mue} by ${\rm MIN}[\vw\, , \,c_l\times{\rm
MIN}\{1,{3\over 2} {c_l\over c_w}\}]$. It turns out that the value
of $\alpha$ used by Binette et\,al. (1999) is a factor 3 too large.
This can be corrected by dividing by 3 the  values of the layer's
thickness, $h$, quoted in their paper.



\clearpage


\clearpage
\end{document}